\documentclass[aps,12pt,final,notitlepage,oneside,onecolumn,
nobibnotes,nofootinbib,superscriptaddress,noshowpacs,centertags,amsmath,amssymb]{revtex4}
\usepackage{epsfig}
\usepackage{bm}
\usepackage[usenames]{color}

\begin{document}

\title{
A unique signal of excited bosons in dijet data from
$\boldsymbol{pp}\,$-collisions}
\author{M. V. Chizhov$^{1,2}$, V. A. Bednyakov$^1$, J. A. Budagov}
\affiliation{Dzhelepov Laboratory of Nuclear Problems,\\
\mbox{Joint Institute for Nuclear Research, 141980, Dubna,
Russia}\\
$^{\it 2}$Centre for Space Research and Technologies, Faculty of
Physics, Sofia University, 1164 Sofia, Bulgaria}


\begin{abstract}
With this note we would like to draw attention to a possible novel
signal of new physics in dijet data at the hadron colliders. Usually
it is accepted that all exotic models predict that these two jets
populate the central (pseudo)rapidity region where $y_{1,2}\simeq
0$. Contrary, the excited bosons do not contribute into this region,
but produce an excess of dijet events over the almost flat QCD
background in $\chi=\exp|y_1-y_2|$ away from this region.
\end{abstract}


\maketitle

\section{Introduction}

Due to the largest cross section of all processes at the hadron
colliders dijet production opens a possibility to search for a
signal of new physics in the very early data. In particular, a bump
in the dijet invariant mass spectrum would indicate the presence of
a resonance decaying into two energetic partons. However, due to the
huge QCD background the bump could be just a statistical fluctuation
of the limited integrated luminosity of the fist data. Besides this,
the bump in the invariant mass distribution stems from the
Breit--Wigner propagator form, which is characteristic for any type
of the resonance regardless of its other properties, like spin,
internal quantum number, etc. Therefore, other observables are
necessary in order to confirm the bump and to reveal the resonance
properties.

The distribution of dijets over
     the polar angle $\theta$\,\footnote{The polar angle $\theta$ is angle between the
axis of the jet pair and the beam direction in the dijet rest
frame~\cite{CS}.} is directly sensitive to the resonance spin and
the dynamics of the underlying process. While the QCD processes are
dominated by $t$-channel gluon exchanges, which lead to a
Rutherford-like distribution $1/(1-\cos\theta)^2$, exotic physics
processes proceed mainly through the $s$-channel, where the spin of
the resonance uniquely defines the angular distribution. For
high-mass resonances and practically massless partons it is
convenient to use the helicity formalism, since the helicity is a
good quantum number for massless particles.

For example, the decay angular distribution in the
centre-of-momentum frame of a particle with spin-$s$ and helicity
$\lambda$ with $-s\leq\lambda\leq s$ decaying into two massless
particles with helicities $\lambda_1$ and $\lambda_2$ can be written
as~\cite{Haber}
\begin{equation}\label{ds/dcos}
    \frac{{\rm d}\Gamma_s}{{\rm d}\cos\theta\;{\rm d}\phi}=
    \frac{1}{64\pi^2 M}\,\vert\,{\cal
    M}^{s\lambda}_{\lambda_1\lambda_2}(\theta,\phi)\vert^2,
\end{equation}
where the helicity amplitude
\begin{equation}\label{amplitude}
    {\cal M}^{s\lambda}_{\lambda_1\lambda_2}(\theta,\phi)=
    \sqrt{\frac{2s+1}{4\pi}}\,e^{i(\lambda-\delta)\phi}\,
    d^s_{\lambda\delta}(\theta)\,{\cal M}^s_{\lambda_1\lambda_2}
    \vspace{0.2cm}
\end{equation}
is expressed through the difference
$\delta\equiv\lambda_1-\lambda_2$ with $-s\leq\delta\leq s$ and the
reduced decay amplitude ${\cal M}^s_{\lambda_1\lambda_2}$, which is
a function of $s$ and the helicities of the outgoing particles, but
is independent on the azimuthal  ($\phi$) and the polar ($\theta$)
angles.
The $\theta$-dependence is concentrated only in the well-known
$d$-functions $d^s_{\lambda\delta}(\theta)$.

The absolute value of the dijet rapidity difference is related to
the polar scattering angle $\theta$ with respect to the beam axis by
$\Delta y\equiv |y_1-y_2|=\ln[(1+|\cos\theta|)/(1-|\cos\theta|)]\ge
0$ and is invariant under boosts along the beam direction. The
choice of the other variable $\chi\equiv\exp(\Delta
y)=(1+|\cos\theta|)/(1-|\cos\theta|)\ge 1$ is motivated by the fact
that Rutherford scattering does not depend on it.

In the next section
we will analyse a model independent signal of new physics using
distributions on these variables and their combinations.

\section{A unique signal of excited bosons}

Let us consider different possibilities for the spin of a resonance
and its possible interactions with partons. The simplest case of the
resonance production of a (pseudo)scalar particle $h$ with spin 0 in
$s$-channel leads to a uniform decay distribution on the scattering
angle
\begin{equation}\label{G0}
    \frac{{\rm d} \Gamma_0(h\to q\bar{q})}
    {{\rm d} \cos\theta} \propto \vert d^{\,0}_{00}\vert\,^2\sim 1.
\end{equation}

The spin-1/2 fermion resonance, like an excited quark $q^*$, leads
to asymmetric decay distributions for the given spin parton
configurations
\begin{equation}\label{G12}
    \frac{{\rm d} \Gamma_{1/2}(q^*\to qg)}
    {{\rm d} \cos\theta} \propto \vert d^{1/2}_{1/2,1/2}\vert\,^2\sim 1+\cos\theta
\end{equation}
and
\begin{equation}\label{G-12}
    \frac{{\rm d} \Gamma_{1/2}(q^*\to qg)}
    {{\rm d} \cos\theta} \propto \vert d^{1/2}_{1/2,-1/2}\vert\,^2\sim
    1-\cos\theta.
\end{equation}
However, the choice of the variables, which depend on the absolute
value of $\cos\theta$, cancels out the apparent dependence on
$\cos\theta$. In other words, the distributions (\ref{G12}) and
(\ref{G-12}) for dijet events look like uniform distributions on
$\Delta y$ and $\chi$.

\begin{figure}[th]
\epsfig{file=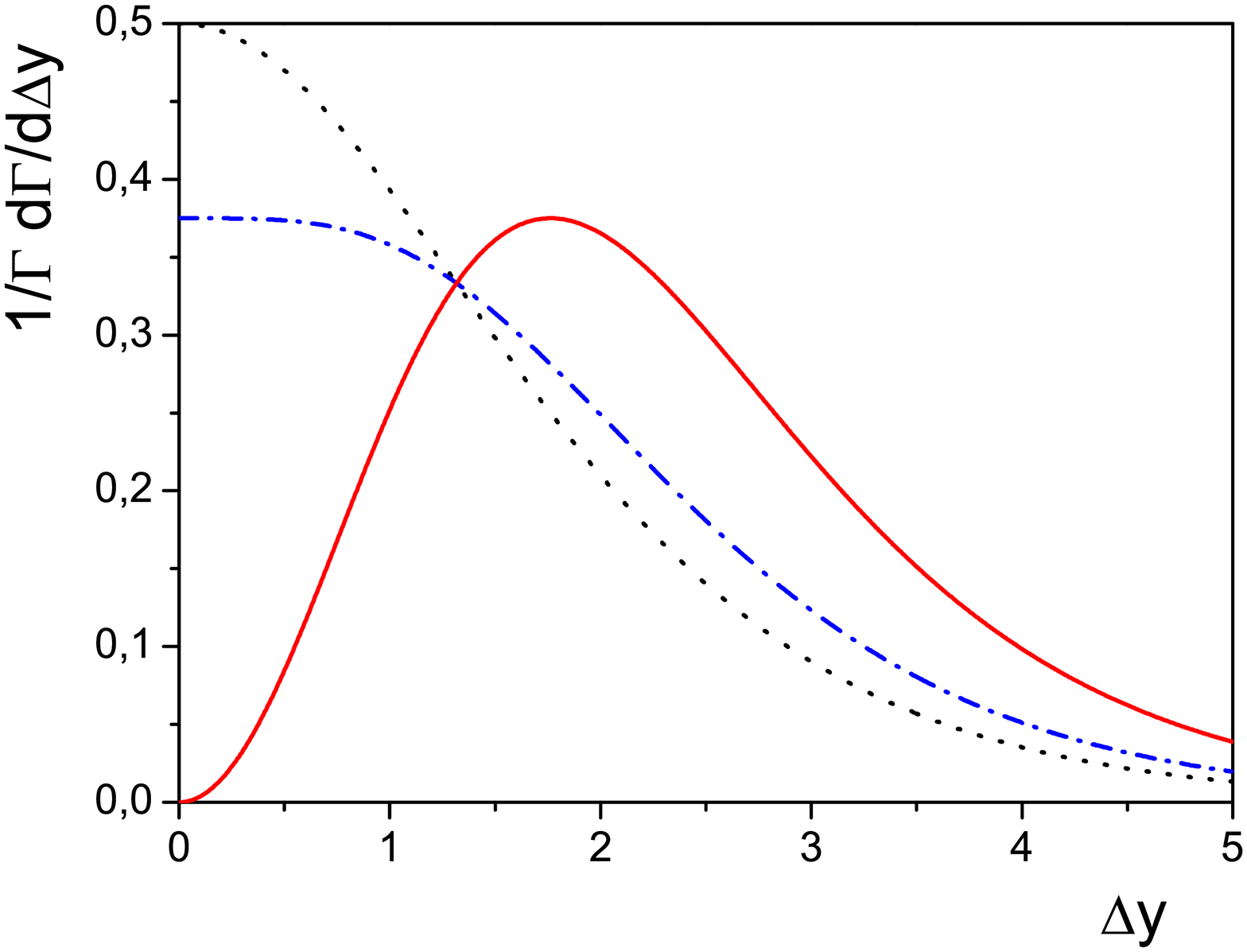,width=0.5\textwidth}\epsfig{file=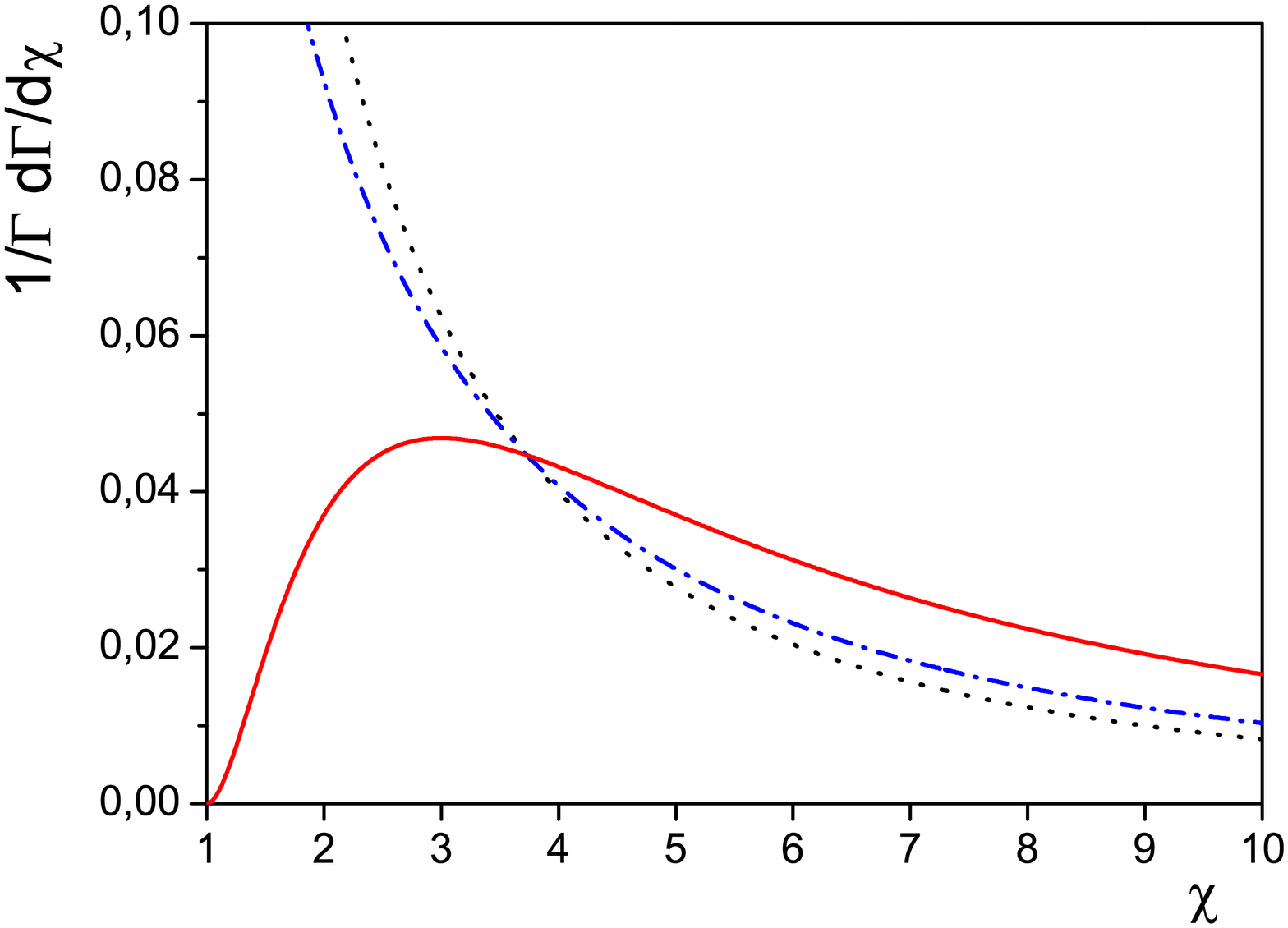,width=0.5\textwidth}
\caption{\label{fig:angular} The normalized angular dijet
distributions as functions of the absolute value of the rapidity
difference (left) and the $\chi$ variable (right) for the scalar
(or/and spin-1/2), the gauge bosons with the minimal coupling and
the excited bosons are shown correspondingly with the dotted,
dash-dotted and solid curves.}
\end{figure}

According to the simple formula
\begin{equation}\label{trans}
    \frac{{\rm d}\Gamma}{{\rm d}(\Delta y/\chi)}=
    \frac{{\rm d}\cos\theta}{{\rm d}(\Delta y/\chi)}\;
    \frac{{\rm d}\Gamma}{{\rm d}\cos\theta}\,,
\end{equation}
the uniform distribution leads to kinematical peaks at small values
of $\Delta y=0$ (the dotted curve in the left panel of
Fig.~\ref{fig:angular})
\begin{equation}\label{dy0}
    \frac{{\rm d}\Gamma_0}{{\rm d}\Delta y}\propto
    \frac{{\rm e}^{\Delta y}}{({\rm e}^{\Delta y}+1)^2}
\end{equation}
and $\chi=1$
(the dotted curve in the right panel of Fig.~\ref{fig:angular})
\begin{equation}\label{chi0}
    \frac{{\rm d}\Gamma_0}{{\rm d}\chi}\propto
    \frac{1}{(\chi+1)^2}.
\end{equation}

A novel situation can occur for the spin-1 resonances, which appear
as new gauge bosons in the extended gauge symmetry groups. We will
consider all possible  interactions of such bosons with the ordinary
fermions. There are two different possibilities.

The gauge bosons, which are associated with additional $U(1)'$ gauge
symmetry (or transform under the adjoint representation of the extra
gauge group), are generally called $Z'(\boldsymbol{W}')$ particles.
They have minimal gauge interactions with the known light fermions
\begin{equation}\label{LL}
    {\cal L}_{Z'}=\sum_f \left(g^f_{LL}\;\overline{\psi^f_L}\gamma^\mu\psi^f_L
    +g^f_{RR}\;\overline{\psi^f_R}\gamma^\mu\psi^f_R\right) Z'_\mu \,
    ,
\end{equation}
which preserve the fermion chiralities and possess maximal
helicities $\lambda=\pm 1$. At a symmetric $pp$ collider, like the
LHC, such interactions lead to the specific symmetric angular
distribution of the resonance decay products over the polar angle
$\theta$,
\begin{equation}\label{GLL}
    \frac{{\rm d} \Gamma_1(Z'\to q\bar{q})}
    {{\rm d} \cos\theta} \propto
    \vert d^1_{11}\vert^2+\vert d^1_{-11}\vert^2 \sim
    1+\cos^2\theta \, .
\end{equation}
Similar to the uniform distribution, eq.~(\ref{GLL}) leads to
kinematical peaks also at small values of $\Delta y=0$  (the
dash-dotted curve in the left panel of Fig.~\ref{fig:angular})
\begin{equation}\label{dy1prime}
    \frac{{\rm d}\Gamma'_1}{{\rm d}\Delta y}\propto
    \frac{{\rm e}^{\Delta y}({\rm e}^{2\Delta y}+1)}{({\rm e}^{\Delta y}+1)^4}
\end{equation}
and $\chi=1$  (the dash-dotted curve in the right panel of
Fig.~\ref{fig:angular})
\begin{equation}
\label{chi1prime}
    \frac{{\rm d}\Gamma'_1}{{\rm d}\chi}\propto
    \frac{\chi^2+1}{(\chi+1)^4}.
\end{equation}

Another possibility is the resonance production and decay of new
longitudinal spin-1 bosons with helicity $\lambda=0$. The new gauge
bosons with such properties arise in many extensions~\cite{Gia} of
the Standard Model (SM), which solve the Hierarchy Problem.
They transform as doublets $(Z^*\;W^*)$ under the fundamental
representation of the SM $SU(2)_W$ group like the SM Higgs boson.

While the $Z'$ bosons with helicities $\lambda=\pm 1$ are produced
in left(right)-handed quark and right(left)-handed antiquark fusion,
the longitudinal $Z^*$ bosons can be produced through the anomalous
chiral couplings with the ordinary light fermions
\begin{equation}\label{LR}
    {\cal L}_{Z^*}=\sum_{f}\left(\frac{g^f_{LR}}{M}\,
    \overline{\psi^f_L}\sigma^{\mu\nu}\psi^f_R\;\partial_{[\mu}
    \bar{Z}^*_{\nu]}
    +\frac{g^f_{RL}}{M}\,\overline{\psi^f_R}\sigma^{\mu\nu}
    \psi^f_L\;\partial_{[\mu} Z^*_{\nu]}\right)
\end{equation}
in left-handed or right-handed quark-antiquark
fusion~\cite{proposal}. The anomalous interactions (\ref{LR}) are
generated on the level of the quantum loop corrections and can be
considered as effective interactions. The gauge doublets, coupled to
the tensor quark currents, are some types of ``excited'' states as
far as the only orbital angular momentum with $L=1$ contributes to
the total angular moment, while the total spin of the system is
zero. This property manifests itself in their derivative couplings
to fermions and a different chiral structure of the interactions in
contrast to the minimal gauge interactions (\ref{LL}).

The anomalous couplings lead to a different angular distribution of
the resonance decay
\begin{equation}\label{GLR}
    \frac{{\rm d} \Gamma^*_1(Z^*\to q\bar{q})}
    {{\rm d} \cos\theta} \propto
    \vert d^1_{00}\vert\,^2\sim\cos^2\theta.
\end{equation}
than the previously considered ones. At first sight, the small
difference between the distributions (\ref{GLL}) and (\ref{GLR})
seems unimportant. However, the absence of the constant term in the
latter case results in novel experimental signatures.

\begin{figure}[h]
\epsfig{file=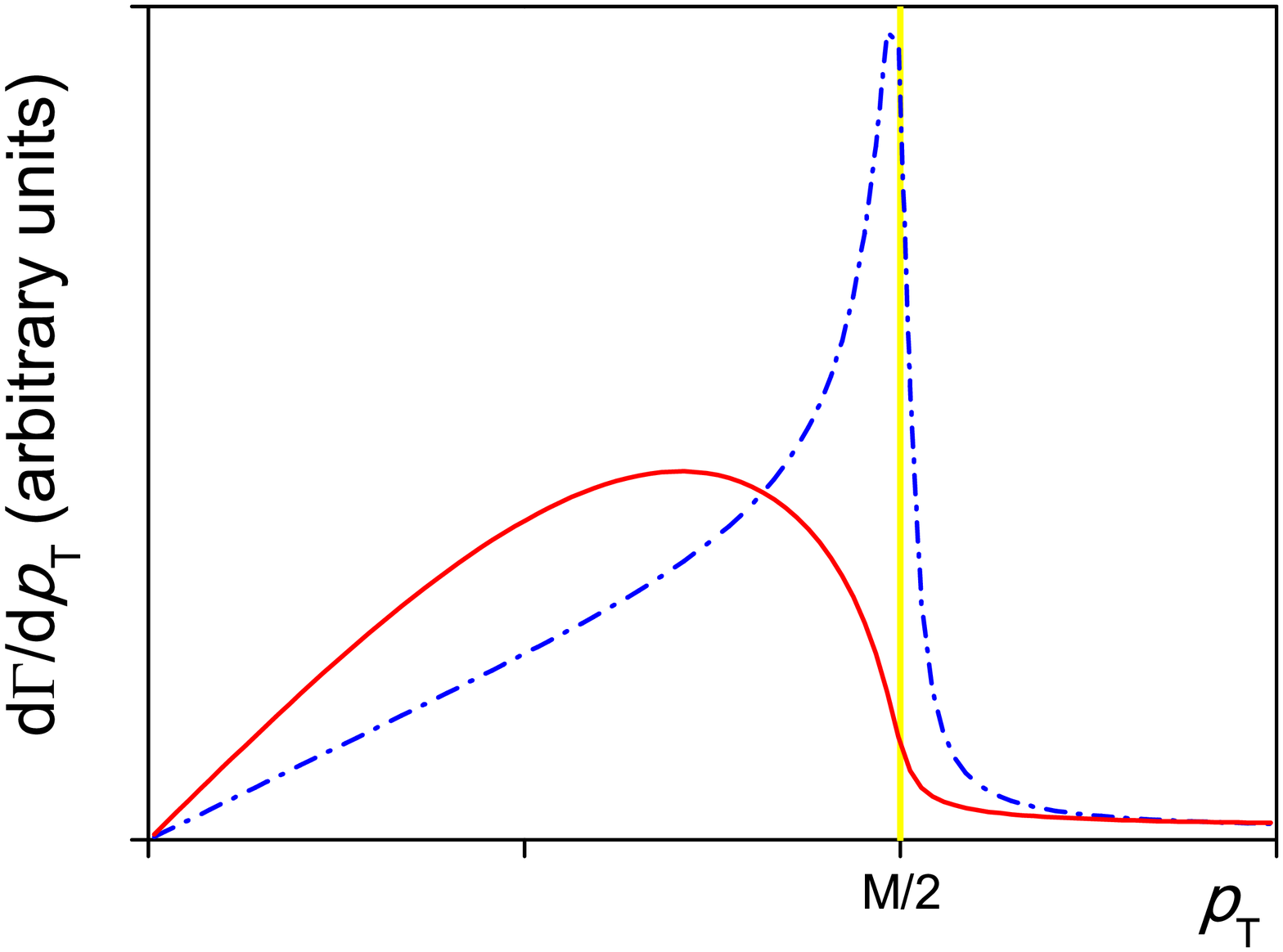,width=0.49\textwidth}\epsfig{file=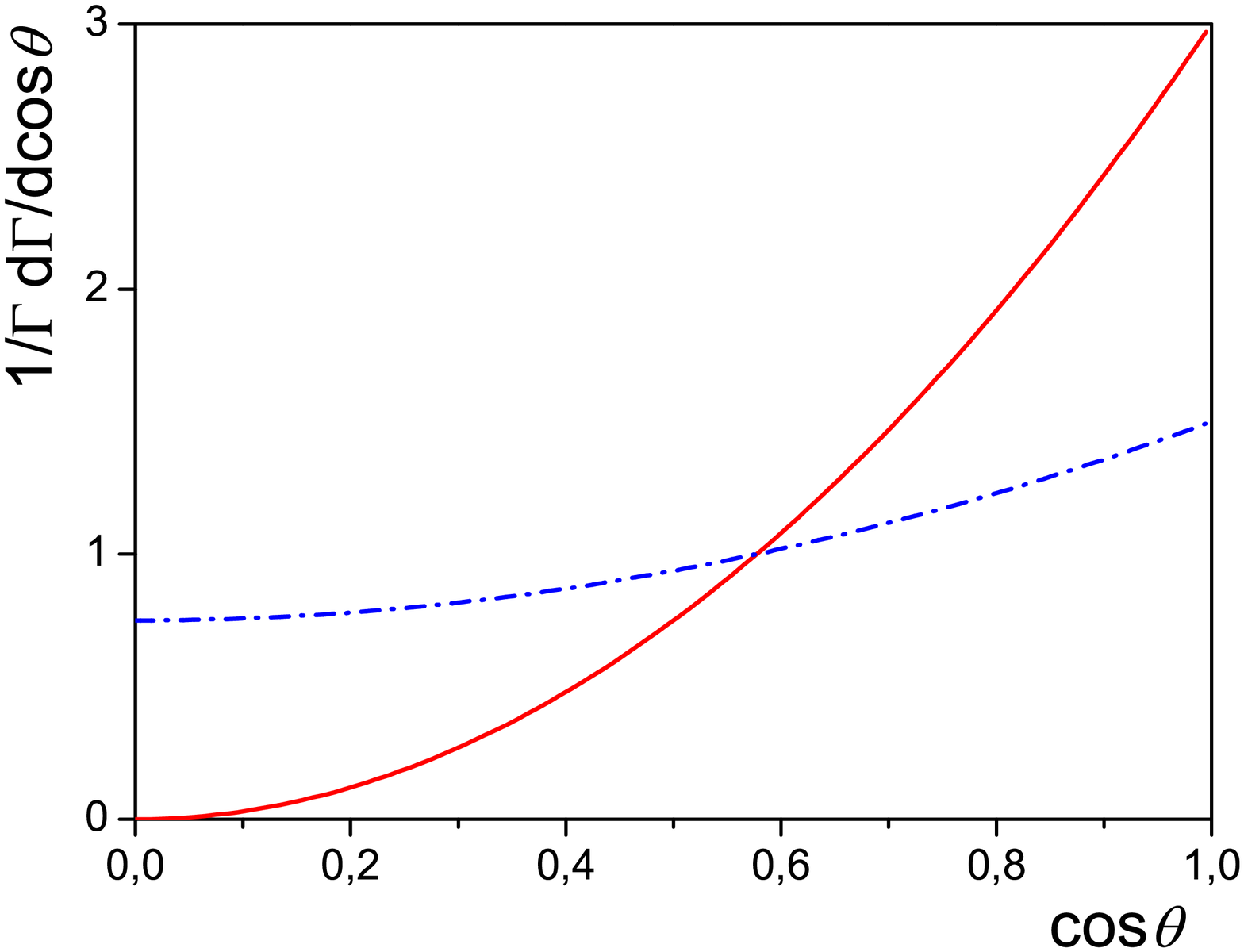,width=0.49\textwidth}
\caption{\label{fig:pt} The final parton transverse momentum (left)
and angular (right) distributions from the gauge (dash-dotted
curves) and the excited (solid curves) bosons decays.}
\end{figure}

First of all, the uniform distribution (\ref{G0}) for scalar and
spin-1/2 particles and the distribution (\ref{GLL}) for gauge vector
bosons with minimal coupling include a nonzero constant term, which
leads, to a kinematic singularity in the transverse momentum
distribution of the final parton
\begin{equation}\label{1/cos}
    \frac{{\rm d}\cos\theta}{{\rm d}p_{\rm T}}\sim
    \frac{1}{\cos\theta}\propto\frac{1}{\sqrt{(M/2)^2-p^2_{\rm T}}}
\end{equation}
in the narrow width approximation $\Gamma \ll M$
\begin{equation}\label{narrow}
    \frac{1}{(s-M^2)^2+M^2\Gamma^2}\approx\frac{\pi}{M\Gamma}\,\delta(s-M^2).
\end{equation}
After smearing from the resonance finite width the singularity is
transformed into the well-known Jacobian peak (the dash-dotted curve
in the left panel of Fig.~\ref{fig:pt}). The analytic expression of
the $p_{\rm T}$ distribution can be found in \cite{Barger}.

Using the same method one can derive the analogous distribution for
the excited bosons (the solid curve in the left panel of
Fig.~\ref{fig:pt}).
\begin{equation}\label{pTanal}
    \frac{{\rm d}\Gamma^*_1}{{\rm d}p_{\rm T}}\propto
    \frac{p_{\rm T}}{\sqrt{\sqrt{\left(4p^2_{\rm T}-M^2\right)^2+\Gamma^2 M^2}+4p^2_{\rm
    T}-M^2}}.
\end{equation}
In contrast to the previous case, the pole in the decay distribution
of the excited bosons is canceled out and the final parton $p_{\rm
T}$ distribution has a broad smooth hump~\cite{ICTP} with a maximum
at $p_{\rm T}=\sqrt{(M^2+\Gamma^2)/8}\simeq M/\sqrt{8}$ below the
kinematical endpoint $p_{\rm T}=M/2$, instead of a sharp Jacobian
peak, that obscures their experimental identification as resonances.
Therefore, the transverse jet momentum is not the appropriate
variable for the excited boson search.

Another striking feature of the distribution (\ref{GLR}) is the
forbidden decay direction perpendicular to the boost of the excited
boson in the rest frame of the latter (the Collins--Soper
frame~\cite{CS}). It leads to a profound dip at $\cos\theta=0$ in
the Collins--Soper frame~\cite{proposal} in comparison with the
gauge boson distribution (the right panel in Fig.~\ref{fig:pt}). The
same dips present also at small values of $\Delta
y=0$~\cite{LaTuile10} (the solid curve in the left panel of
Fig.~\ref{fig:angular})
\begin{equation}\label{dy1star}
    \frac{{\rm d}\Gamma^*_1}{{\rm d}\Delta y}\propto
    \frac{{\rm e}^{\Delta y}({\rm e}^{\Delta y}-1)^2}{({\rm e}^{\Delta y}+1)^4}
\end{equation}
and $\chi=1$ (the solid curve in the right panel of
Fig.~\ref{fig:angular})
\begin{equation}\label{chi1star}
    \frac{{\rm d}\Gamma^*_1}{{\rm d}\chi}\propto
    \frac{(\chi-1)^2}{(\chi+1)^4}.
\end{equation}

It can be seen from Fig.~\ref{fig:angular} that the excited bosons
have unique signature in the angular distributions. They manifest
themselves through the absolute minima at small values of $\Delta
y=0$ and $\chi=1$ and absolute maxima right away from the origin.
So, the rapidity difference distribution reaches the absolute
maximum at $\Delta y=\ln(3+\sqrt{8})\approx 1.76$ and at $\chi=3$
for the angular distribution on the dijet variable $\chi$.

\section{First look at preliminary LHC data}
\begin{figure}[th]
\epsfig{file=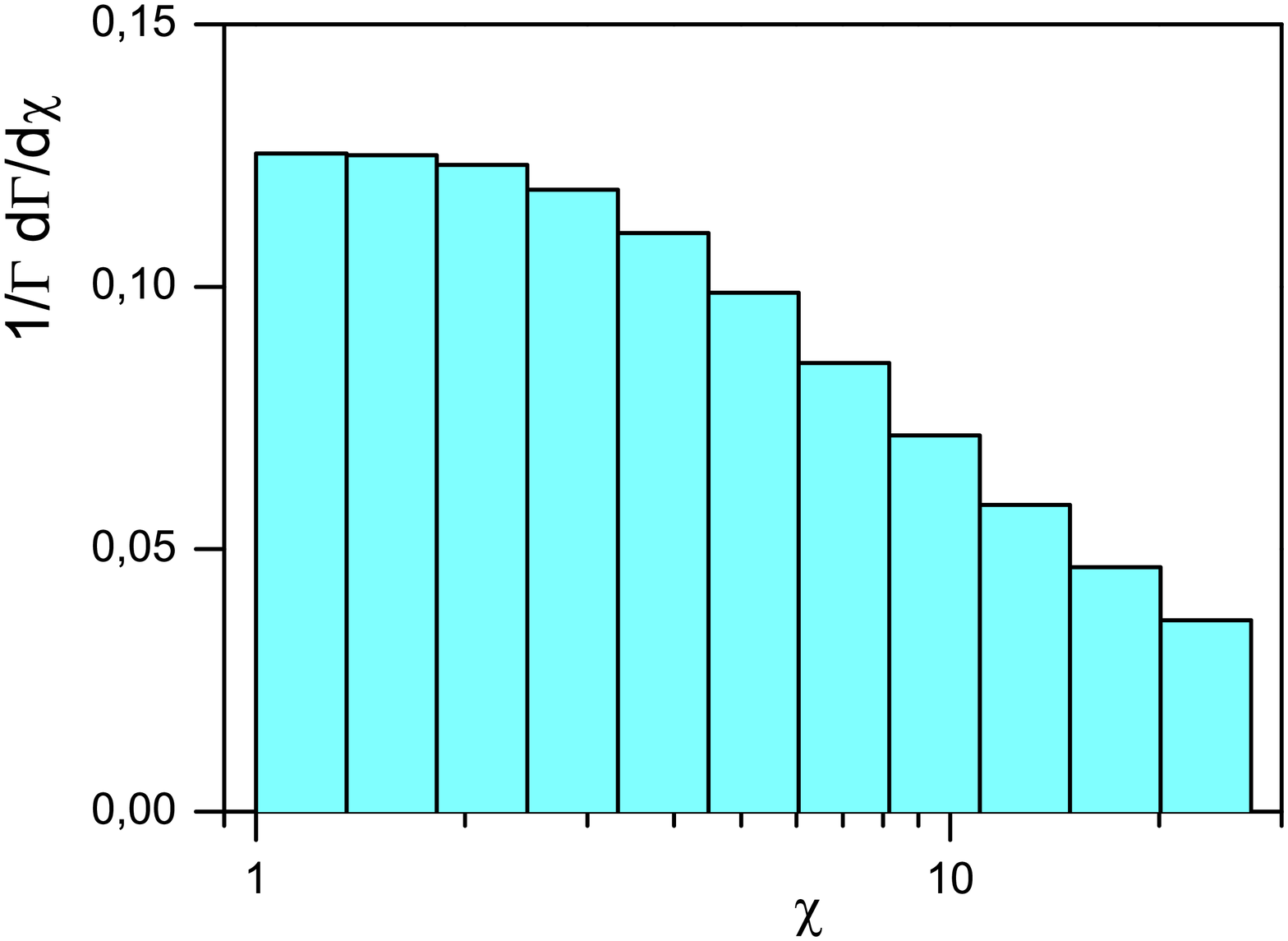,width=0.5\textwidth}\epsfig{file=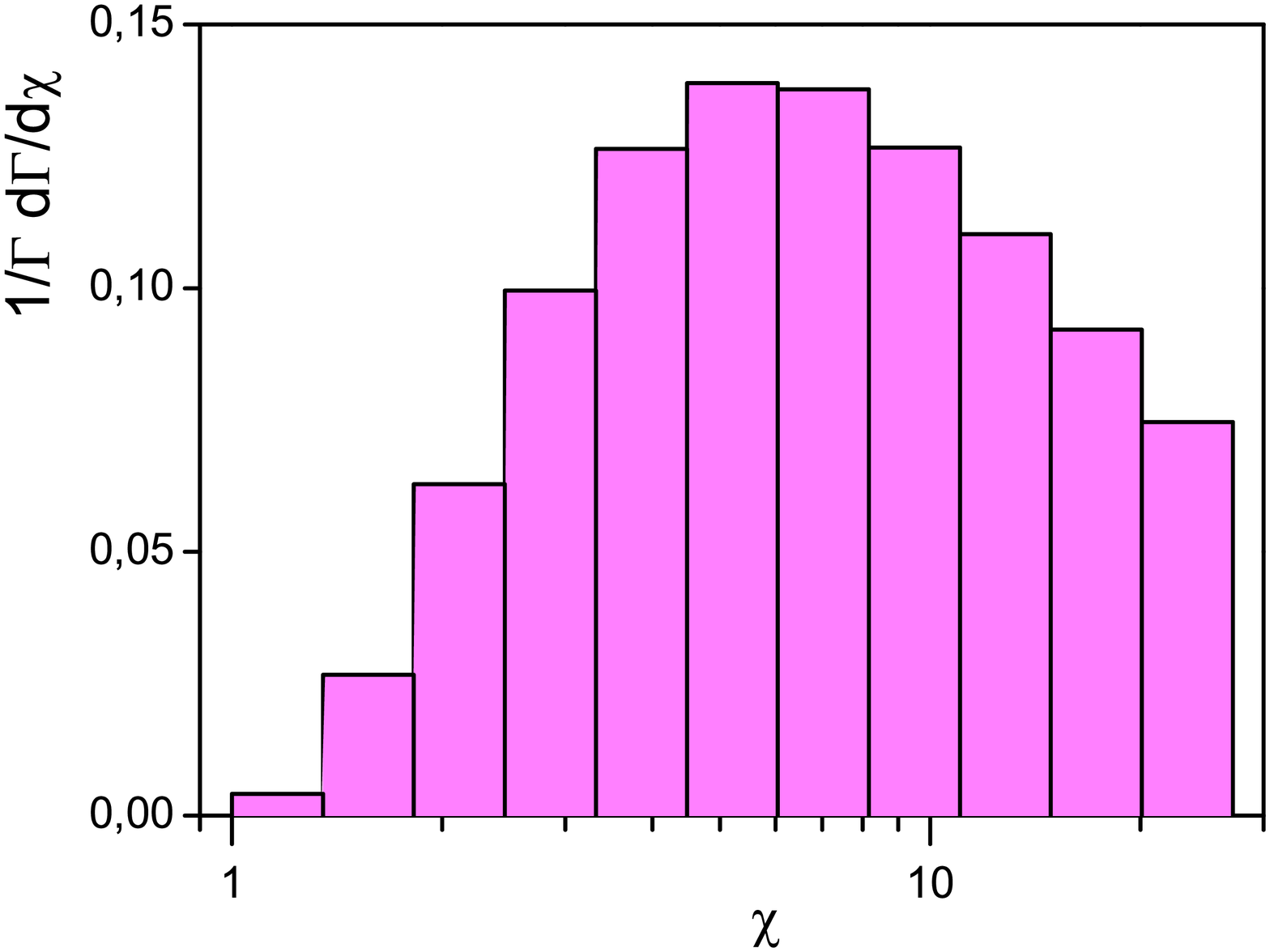,width=0.5\textwidth}
\caption{\label{fig:chi} The normalized histograms of $\chi$-spectra
for the gauge bosons with the minimal coupling (left) and for the
excited bosons (right).}
\end{figure}

In order to satisfy to bin purity and stability for physics studies
it is convenient to use equidistant binning in
$\log\chi$~\cite{PLB}, which corresponds to periodic cell
granularity of calorimeter in $\eta$. In this case the smooth
$\chi$-spectra (see eqs. (\ref{chi1prime}) and (\ref{chi1star}))
transform into histograms with maximum in the lowest bin for the
gauge bosons with the minimal coupling and with maximum in the bin
containing value of $\chi=3+\sqrt{8}\approx 5.8$ for the excited
bosons (Fig.~\ref{fig:chi}).
It is interesting to note that a shape similar to the excited-boson
one (right panel in Fig.~\ref{fig:chi}) can be observed in the
preliminary ATLAS data~\cite{PLB} in $\chi$-distribution for the
highest dijet mass region $M_{jj}>1200$~GeV.

Using distributions in (pseudo)rapidity and $\chi$ we can construct
two useful ratios: the wide-angle to small-angle ratio
\begin{equation}\label{wide-small}
    R_\chi(a,b)=\frac{N(1<\chi<a)}{N(a<\chi<b)}
\end{equation}
and the centrality or $\eta$-ratio of both jets
\begin{equation}\label{centrality}
    R_\eta(a,b)=\frac{N(|\eta_{1,2}|<a)}{N(a<|\eta_{1,2}|<b)},
\end{equation}
which are less affected by the systematic errors and, therefore,
used for search of new physics in dijet data.

Let us suppose that we have found some bump in the dijet invariant
mass distribution. For example, an excess at 550~GeV can be seen in
the preliminary ATLAS data~\cite{bumpATLAS} for dijet processes at
0.315 nb$^{-1}$ of integrated luminosity.\footnote{However, updated
results~\cite{newATLAS}, based on 3.1 pb$^{-1}$, do not confirm this
excess.} Then we can compare the angular distributions for ``on
peak'' and ``off peak'' events, using, for example, the
aforementioned ratios.\footnote{We ignore the complications of
experimental separation of ``on peak'' events from the ``off peak''
ones.} Since the QCD backgrounds is dominated by the Rutherford-like
distribution, we can consider, as an approximation, a simple case
when the QCD dijet $\chi$-distribution is flat. It means that for
the selected equal regions $b-a=a-1$ ($b=2a-1$), the $R_\chi$ ratio
for the ``off peak'' events should be approximately one and does not
depend on the dijet mass.

In the case, when the ``on peak'' events have originated from the
new physics and have angular distribution different from the QCD
one, this $R_\chi$ ratio should deviate from one. All known exotic
models, besides the excited bosons, predict $R_\chi>1$ due to an
excess at low $\chi$ values independent on $a$. In order to
emphasize the effect of new physics in the case of the specific
signature of the excited bosons, when $R_\chi<1$, we need to choose
$a$ value, which minimizes the ratio
\begin{equation}\label{minChi}
R_\chi(a,2a-1)=
\frac{N_{\rm QCD}+N_{\rm new}({\rm low-}\chi)}
     {N_{\rm QCD}+N_{\rm new}({\rm high-}\chi)}
\approx 1+
\frac{N_{\rm new}({\rm low-}\chi)
    -N_{\rm new}({\rm high-}\chi)}{N_{\rm QCD}}<1.
\end{equation}
Since $N_{\rm QCD}\propto (a-1)$, $N_{\rm new}({\rm low-}\chi)
\propto (a-1)^3/(a+1)^3$ and $N_{\rm new}({\rm high-}\chi) \propto
(a-1)^3/a^3-N_{\rm new}({\rm low-}\chi)$, it is possible for the
choice of the parameters $a\approx 1.87$ and $b\approx 2.74$ due to
the monotonic increase of the distribution for the excited bosons
until the maximum at $\chi=3$. The larger value of $a\ge
1/(\sqrt[3]{2}-1)\approx 3.85$ will lead to a compensation of the
contributions from low and high $\chi$-parts and $R_\chi \ge 1$.

For the fixed parameters $a$ and $b$ the centrality ratio $R_\eta$
for the QCD processes is also almost constant and should not depend
on the dijet invariant mass. In the case of a presence of new
physics signal this ratio could deviate from its constant value. For
almost all exotic models the signal events are concentrated in the
central region and usually this lead to a bump in $R_\eta$
distribution versus the dijet masses. By contrast to this, the
signal from the excited bosons can lead to a novel signature: a dip
in the distribution at the resonance mass.

\begin{figure}[th]
\hspace{-0.5cm}\epsfig{file=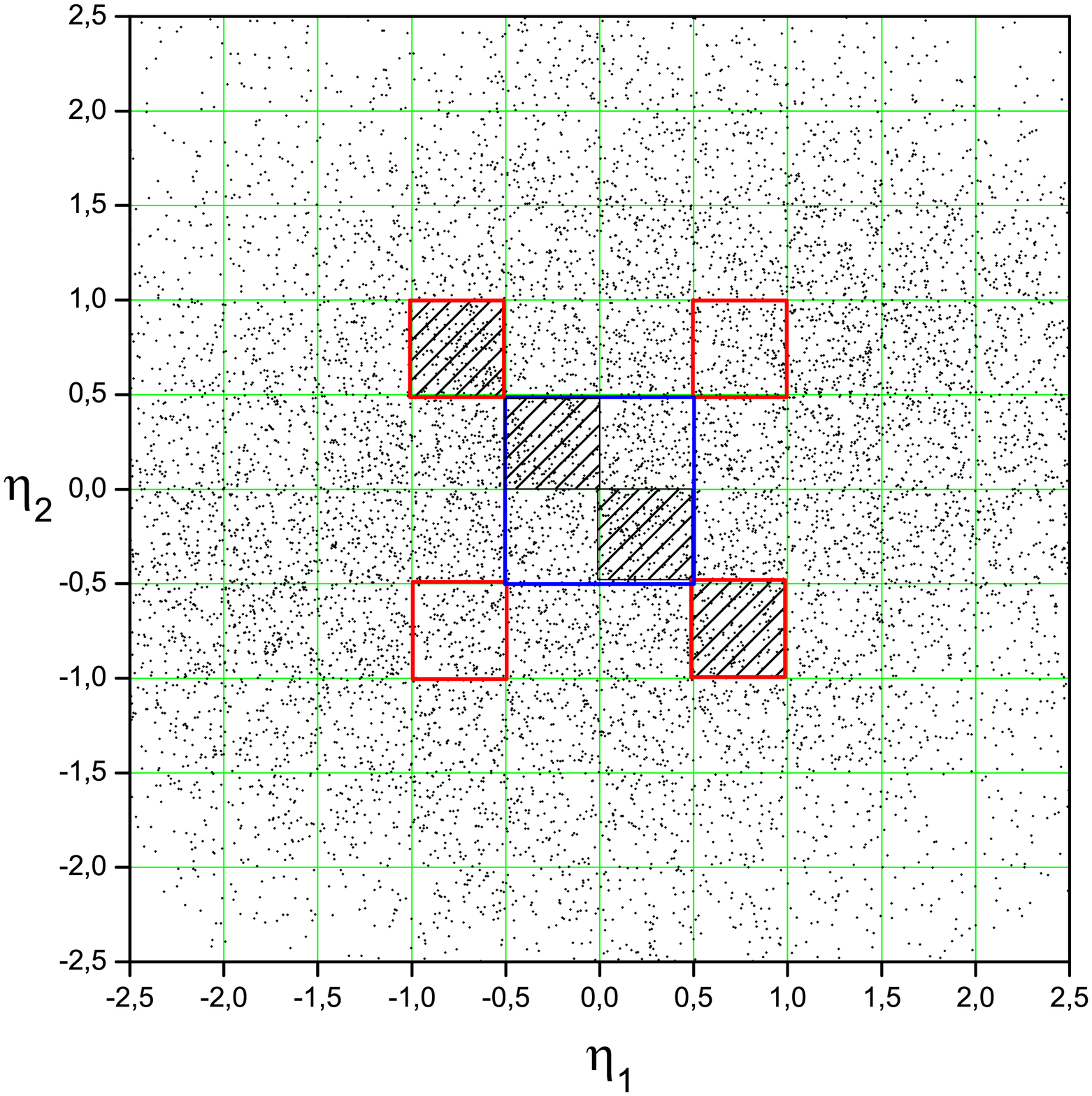,width=0.55\textwidth}
\hspace{-1.5cm}\epsfig{file=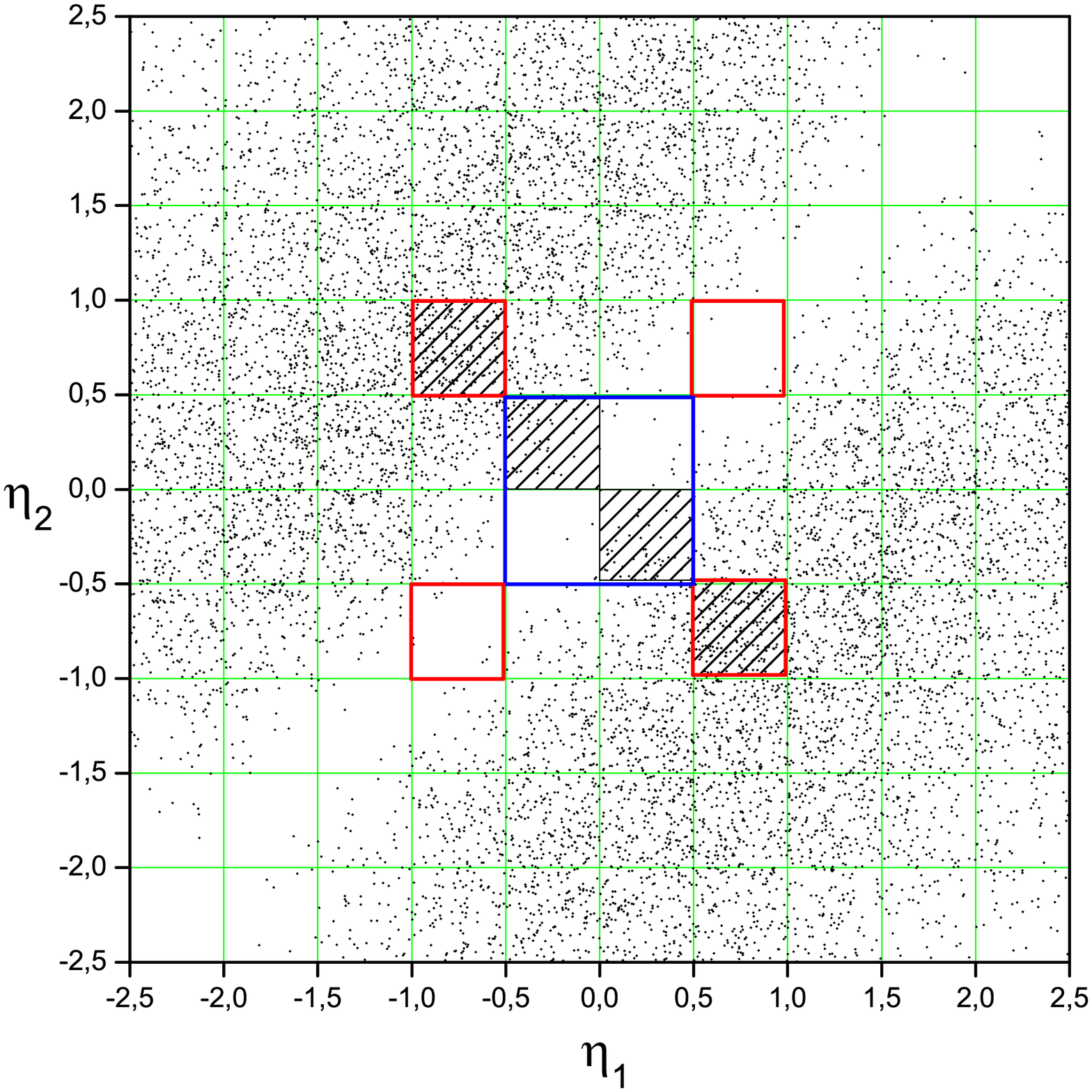,width=0.55\textwidth}
\caption{\label{fig:eta2dim} The two dimensional pseudorapidity
distributions (scatter plots) of dijet events for the gauge bosons
with the minimal coupling (left) and for the excited bosons (right).
The central region $|\eta_{1,2}|<0.5$ and outer regions
$0.5<|\eta_{1,2}|<1.0$ are depicted. The hatched squared regions
correspond to selected events with opposite pseudorapidity signs.}
\end{figure}

We have used the CompHEP package~\cite{CompHEP} and the extended
model with the excited bosons~\cite{EMS} to generate two dimensional
pseudorapidity distributions for ``$\,2\to 2\,$'' processes
proceeding through the gauge $W'$ and excited $W^*$ boson resonances
with the same mass 550 GeV at $\sqrt{s}=7$~TeV $pp$ collider
(Fig.~\ref{fig:eta2dim}).
The CTEQ6L parton distribution functions were used. For both final
jets we impose cuts on the pseudorapidity $|\eta|<2.5$ and the
transverse momentum $p_{\rm T} > 30$~GeV.

To minimize the potential differences in jet response between the
{\em inner} and {\em outer} dijet events one can choose the central
region of the calorimeter $|\eta|<1$. Using the scatter plots in
Fig.~\ref{fig:eta2dim} one can estimate the centrality ratios
(\ref{centrality}) for the gauge
\begin{equation}\label{etaRprime}
    R'_\eta(0.5,1.0)\simeq 1.08
\end{equation}
and the excited
\begin{equation}\label{etaRstar}
    R^*_\eta(0.5,1.0)\simeq 0.29
\end{equation}
bosons. There is dramatic difference between these two cases, which
should lead to corresponding experimental signature. Since the QCD
ratio $R^{\rm QCD}_\eta(0.5,1.0)\simeq 0.6$ is between the numbers
in (\ref{etaRprime}) and (\ref{etaRstar}) the gauge bosons with the
minimal coupling will lead to increasing the QCD ratio at the
resonance mass, while the excited bosons should decrease the ratio.
Just the second
type of the novel signature can be observed in the distribution of
the $\eta$-ratio versus the dijet invariant mass in the ATLAS
data~\cite{ICHEP10ATLAS} in approximately the same mass range,
$450~{\rm GeV}<M_{jj}<600$~GeV, as for the resonance bump in the
dijet events~\cite{bumpATLAS}.
%

Let us stress here that the extensions of the signal region up to
$|\eta|<1.3$ and the central region up to $|\eta|<0.7$ do not change
drastically the QCD ratio $R^{\rm QCD}_\eta(0.7,1.3)\simeq 0.55$,
but dilute the signal from the excited bosons, since
$R^*_\eta(0.7,1.3)\simeq 0.68$.

In order to increase the sensitivity to new physics one can consider
the centrality ratio only for the dijet events with opposite
pseudorapidities $\left.R_\eta\right|_{(\eta_1\cdot\eta_2)\le
0}={\cal R}_\eta$ (the hatched regions in Fig.~\ref{fig:eta2dim}).
In this case we get a bit larger deviation form one:
\begin{equation}\label{etaR0prime}
    {\cal R}'_\eta(0.5,1.0)\simeq 1.12
\end{equation}
for the gauge and
\begin{equation}\label{etaR0star}
    {\cal R}^*_\eta(0.5,1.0)\simeq 0.25
\end{equation}
for the excited bosons. But in this case we lose half of the
statistics. Therefore, it is convenient to consider the distribution
on $\Delta\eta\equiv|\eta_1-\eta_2|\ge0$ for the events in the
rectangle region $\Delta\eta<b$ and
$\eta_B\equiv|\eta_1+\eta_2|<c$.\footnote{The cut $\eta_B<c$ is
necessarily to reduce an effect of the parton distribution functions
on different $\Delta\eta$ bins.} The corresponding centrality ratio
$R_{\Delta\eta}$ is defined as
\begin{equation}\label{deltaEtaR}
    R_{\Delta\eta}(a,b,c)=\left.\frac{N(\Delta\eta<a)}{N(a<\Delta\eta<b)}\right|_{\eta_B<c}.
\end{equation}

\begin{figure}[th]
\epsfig{file=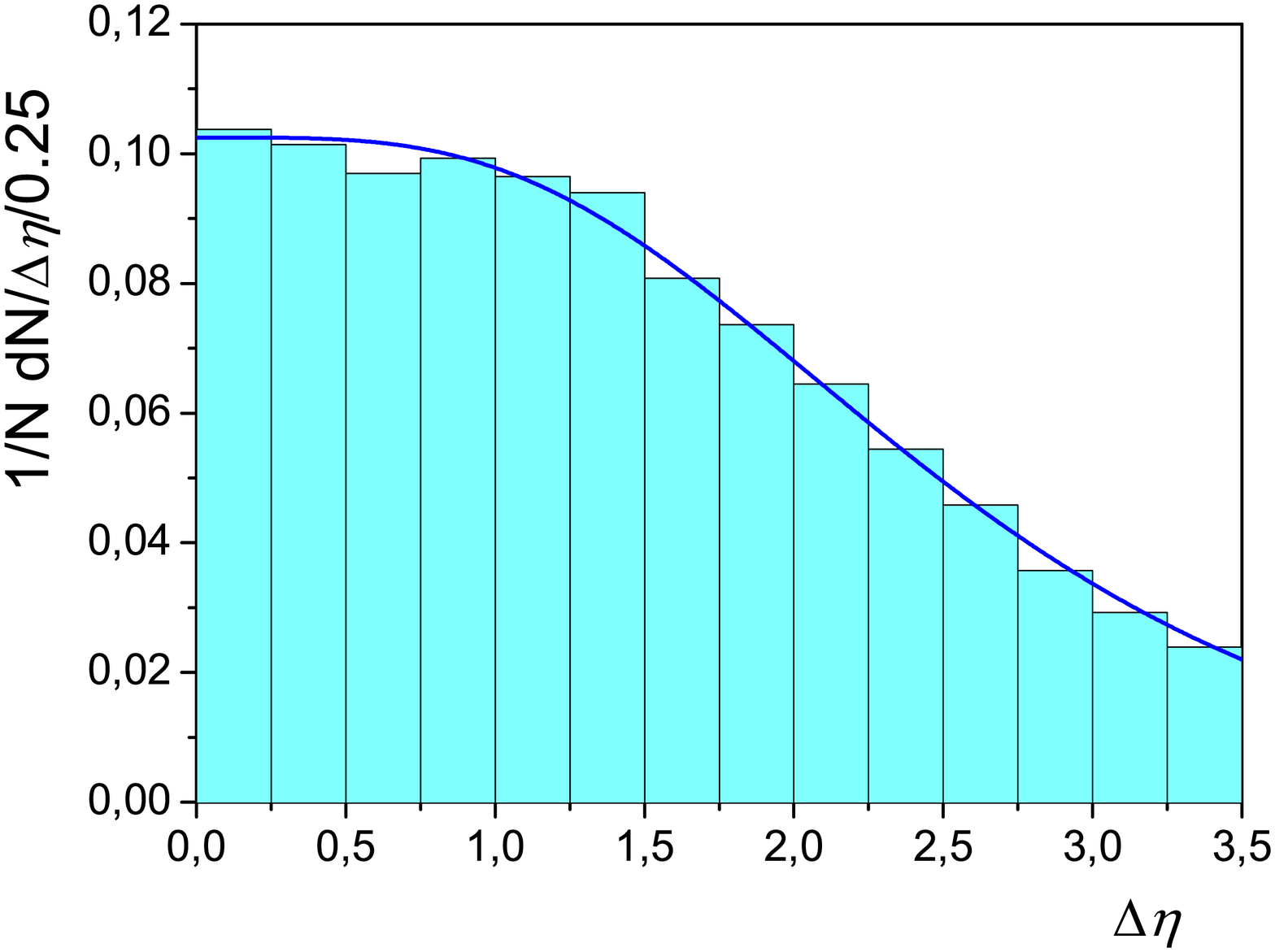,width=0.5\textwidth}\epsfig{file=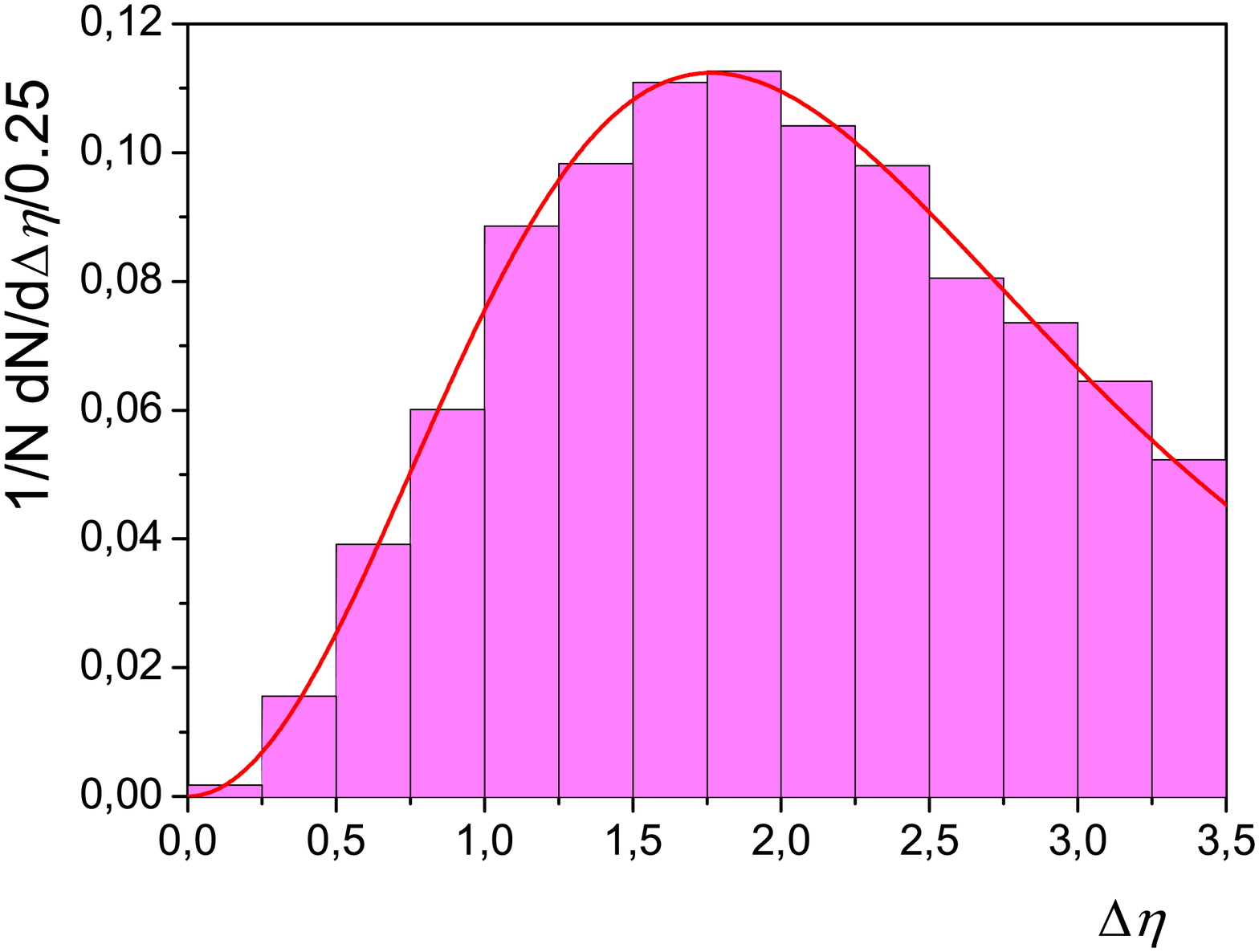,width=0.5\textwidth}
\caption{\label{fig:eta} The normalized histograms of
$\Delta\eta\,$-spectra for the gauge bosons with the minimal
coupling (left) and for the excited bosons (right). The solid curves
correspond to the theoretical formulae (\ref{dy1prime}) and
(\ref{dy1star}).}
\end{figure}

The normalized histograms of $\Delta\eta$-spectra and the
theoretical curves are shown in Fig.~\ref{fig:eta} for the following
parameters values: $b=3.5$ and $c=1.5$.
It can be seen from the figure, that the theoretical distributions
describe very well the above-mentioned simulation data. In the same
way as it was done for the $\chi$-distribution of the excited bosons
one can maximize the deviation from the QCD ratio $R^{\rm
QCD}_{\Delta\eta}=N_{\rm QCD}^{(a)}/N_{\rm QCD}^{(b-a)}$
\begin{equation}\label{minEta}
R_{\Delta\eta}(a,b,c)=
\frac{N_{\rm QCD}^{(a)}+N_{\rm new}^{(a)}}{N_{\rm QCD}^{(b-a)}+N_{\rm new}^{(b-a)}}
\approx R^{\rm QCD}_{\Delta\eta}
    \left(1+\frac{N_{\rm new}^{(a)}}{N_{\rm QCD}^{(a)}}
           -\frac{N_{\rm new}^{(b-a)}}{N_{\rm QCD}^{(b-a)}}
    \right)<1.
\end{equation}
Since $N_{\rm QCD}^{(a)}\propto{\rm e}^{a}-1$, $N_{\rm
new}^{(a)}\propto ({\rm e}^{a}-1)^3/({\rm e}^{a}+1)^3$ and $N_{\rm
QCD}^{(b-a)}\propto{\rm e}^{b}-{\rm e}^{a}$, $N_{\rm
new}^{(b-a)}\propto ({\rm e}^{b}-1)^3/({\rm e}^{b}+1)^3-N_{\rm
new}^{(a)}$, for the central calorimeter $b=1$ the minimum appears
at $a=0.67$\,.

\section{Conclusions}
In this paper we have considered the novel experimental signatures
of the chiral excited bosons in the dijet data. They possess
drastically different angular distributions from all previously
discussed exotic models. For the special choice of parameters the
wide-angle to small-angle ratio and the centrality ratio could be
less than their QCD values, that will be definitely pointing out to
the presence of excited bosons.

\section*{Acknowledgements}
The work of M.V. Chizhov was partially supported by the
JINR-Bulgaria grant for 2010 year.



\begin{thebibliography}{99}
\bibitem{CS} J.C. Collins and D.E. Soper, Phys. Rev. D {\bf 16} (1977) 2219.
\bibitem{Haber} H.E. Haber, ``Spin formalism and applications to new physics searches'',
hep-ph/9405376.
\bibitem{Gia} M.V. Chizhov and Gia Dvali, ``Origin and Phenomenology of Weak-Doublet
Spin-1 Bosons'', arXiv:0908.0924~[hep-ph].
\bibitem{Barger} V. Barger, A.D. Martin and R.J.N. Phillips, Z.
Phys. C {\bf 21} (1983) 99.
\bibitem{ICTP} M.V. Chizhov, ``Production of new chiral bosons at Tevatron and LHC'',
hep-ph/0609141.
\bibitem{proposal} M.V. Chizhov, V.A. Bednyakov and J.A. Budagov,
Phys. Atom. Nucl. {\bf 71} (2008) 2096, arXiv:0801.4235~[hep-ph].
\bibitem{LaTuile10} M.V. Chizhov, V.A. Bednyakov and J.A. Budagov,
``Anomalously interacting extra neutral bosons'',
arXiv:1005.2728~[hep-ph] (to be published in Nuovo Cimento B).
\bibitem{PLB} The ATLAS Collaboration, ``Search for Quark Contact
Interactions in Dijet Angular Distributions in $pp$ Collisions at
$\sqrt{s}=7$~TeV Measured with the ATLAS Detector'',
arXiv:1009.5069~[hep-ex].
\bibitem{bumpATLAS} G. Aad {\em et al.} (ATLAS Collaboration) ``Search for New Particles
in Two-Jet Final States in 7~TeV Proton-Proton Collisions with the
ATLAS Detector at the LHC'', Phys. Rev. Lett. {\bf 105} (2010)
161801, arXiv:1008.2461~[hep-ex].
\bibitem{newATLAS} The ATLAS Collaboration, ``Update of the search for new particles
decaying into dijets in proton-proton collisions at $\sqrt{s}=7$~TeV
with the ATLAS detector'', ATL-CONF-2010-088.
\bibitem{CompHEP} E. Boos {\it et al.} (CompHEP Coll.),
NIM A {\bf 534} (2004) 250, hep-ph/0403113; A.~Pukhov {\it et al.},
INP MSU report 98-41/542, hep-ph/9908288; Home page:
http://comphep.sinp.msu.ru.
\bibitem{EMS} M.V. Chizhov, ``A Reference Model for Anomalously Interacting Bosons'',
arXiv:1005.4287 [hep-ph].
\bibitem{ICHEP10ATLAS} The ATLAS Collaboration, ``High-$p_{\rm T}$ dijet angular
distributions in $pp$ interactions at $\sqrt{s}=7$~TeV measured with
the ATLAS detector at the LHC'', ATL-CONF-2010-074.
\end{thebibliography}
\end{document}